\documentclass[runningheads]{llncs}
\usepackage[T1]{fontenc}
\usepackage{graphicx}
\usepackage{comment}
\usepackage{longtable}
\usepackage{amsmath}
\usepackage{rotating}
\usepackage{etoolbox}
\usepackage{setspace}

\makeatletter
\renewcommand{\@author}{}
\renewcommand{\@institute}{}
\makeatother

\begin{document}
\title{Systematic FAIRness Assessment of Open Voice Biomarker Datasets for Mental Health and Neurodegenerative Diseases}
\titlerunning{Systematic FAIRness Assessment of Open Voice Biomarker Datasets}
% If the paper title is too long for the running head, you can set
% an abbreviated paper title here
%

\author{Ishaan Mahapatra\inst{1}\orcidID{0009-0001-6110-4108} \and \\
Nihar R. Mahapatra\inst{2}\orcidID{0000-0002-3821-3330}}
\authorrunning{I. Mahapatra and N.R. Mahapatra}
% First names are abbreviated in the running head.
% If there are more than two authors, 'et al.' is used.
%
\institute{Haslett High School, Haslett, MI, USA \\ \email{ishaan.mahapatra@gmail.com} \and
Electrical \& Computer Eng., Michigan State University, East Lansing, MI, USA \\
\email{nrm@egr.msu.edu}}
\maketitle              % typeset the header of the contribution
%
%\vspace*{-0.2in}

\begin{abstract}
\textit{Voice biomarkers}—human-generated acoustic signals such as speech, coughing, and breathing—are promising tools for scalable, non-invasive detection and monitoring of mental health and neurodegenerative diseases. Yet, their clinical adoption remains constrained by inconsistent quality and limited usability of publicly available datasets. To address this gap, we present the first systematic FAIR (Findable, Accessible, Interoperable, Reusable) evaluation of 27 publicly available voice biomarker datasets focused on these disease areas. Using the FAIR Data Maturity Model and a structured, priority-weighted scoring method, we assessed FAIRness at subprinciple, principle, and composite levels. Our analysis revealed consistently high Findability but substantial variability and weaknesses in Accessibility, Interoperability, and Reusability. Mental health datasets exhibited greater variability in FAIR scores, while neurodegenerative datasets were slightly more consistent. Repository choice also significantly influenced FAIRness scores. To enhance dataset quality and clinical utility, we recommend adopting structured, domain-specific metadata standards, prioritizing FAIR-compliant repositories, and routinely applying structured FAIR evaluation frameworks. These findings provide actionable guidance to improve dataset interoperability and reuse, thereby accelerating the clinical translation of voice biomarker technologies.
%\vspace*{-0.1in}
\end{abstract}
%
%
%
%\vspace*{-0.2in}
\section{Introduction and Related Work}
\label{sec:intro}

\textit{Voice biomarkers} are human-generated acoustic signals—such as speech, coughing, breathing, crying, and other externally audible vocalizations—that reflect physiological and psychological health status~\cite{fagherazziVoiceHealthUse2021,robin2020evaluation}. Advances in speech signal processing, machine learning, and mobile health technologies have fueled interest in these biomarkers for scalable, non-invasive diagnostics, disease monitoring, and health screening~\cite{fagherazziVoiceHealthUse2021,robin2020evaluation,cummins2018speech}. Voice biomarkers show particular promise for early detection and longitudinal assessment of mental health conditions (e.g., depression, anxiety) and neurodegenerative diseases (e.g., Parkinson’s, Alzheimer’s)~\cite{cummins2018speech,chen2024review,low2020automated}. Their minimally invasive nature, ease of collection using ubiquitous digital devices, and capacity for real-time assessment make them attractive for both clinical and research use.

However, the advancement of voice biomarkers into clinical workflows is hindered by major limitations in publicly available datasets~\cite{robin2020evaluation,clarkAIreadinessBiomedicalData2024}. Common challenges include inconsistent metadata, varied or poorly documented recording protocols, unclear licensing terms, and limited interoperability~\cite{clarkAIreadinessBiomedicalData2024,cummins2018speech,bahimFAIRDataMaturity2020}. These issues obstruct validation, reproducibility, and AI-readiness, leading to 
duplicated research efforts, fragmented knowledge bases, inefficient use of resources, 
and delayed clinical adoption~\cite{clarkAIreadinessBiomedicalData2024,bahimFAIRDataMaturity2020}. Consequently, ensuring the usability, transparency, and interoperability of datasets is essential for translating voice biomarker research into scalable clinical tools.

The FAIR (Findable, Accessible, Interoperable, Reusable) principles offer a structured and widely accepted framework to improve data quality and usability~\cite{wilkinson2016fair,bahimFAIRDataMaturity2020}. Initially proposed by Wilkinson et al.~\cite{wilkinson2016fair}, the FAIR principles promote discoverability, accessibility, metadata standardization, and reuse across domains. Subsequent studies have shown that adopting FAIR principles enhances dataset longevity, research reproducibility, and data-driven innovation~\cite{groth2020fair,bahimFAIRDataMaturity2020,devarajuAutomatedSolutionMeasuring2021}. Applying these principles to voice biomarker datasets offers a clear path to addressing current limitations in data sharing, transparency, and reusability~\cite{clarkAIreadinessBiomedicalData2024,groth2020fair}.

Prior FAIRness evaluations have been conducted across various scientific domains, including genomics, agriculture, and biomedical repositories~\cite{bahimFAIRDataMaturity2020,stellmachHowAssessFAIRness2024,petrosyanFAIRDegreeAssessment2023}. These studies often leverage structured frameworks such as the FAIR Data Maturity Model~\cite{bahimFAIRDataMaturity2020}, and use automated tools like F-UJI, FAIR Evaluator, and FAIR Checker~\cite{devarajuAutomatedSolutionMeasuring2021,stellmachHowAssessFAIRness2024,wilkinson2019evaluating,gaignard2023fair}. While such tools are valuable for scalable evaluation, limitations remain: domain-agnostic metrics can obscure meaningful differences, and partial FAIRness fulfillment is often underrepresented~\cite{devarajuAutomatedSolutionMeasuring2021,stellmachHowAssessFAIRness2024}. For example, we observed that the F-UJI tool often assigns identical FAIRness scores to datasets solely based on shared repositories, despite differences in metadata completeness and accessibility.

Despite growing attention to FAIR principles, no prior work has systematically assessed the FAIRness of publicly available voice biomarker datasets, particularly those targeting mental health and neurodegenerative conditions. Given the clinical relevance and data-specific challenges in this field, such an evaluation is urgently needed to guide data creators, curators, and users toward more interoperable and reusable datasets.

To address this gap, this study presents the first systematic FAIRness evaluation of 27 publicly available voice biomarker datasets focused on mental health and neurodegenerative diseases. Using the FAIR Data Maturity Model~\cite{bahimFAIRDataMaturity2020}, we introduce a novel subprinciple-level scoring methodology that accounts for both full and partial fulfillment of FAIR indicators and incorporates priority weighting to reflect the relative importance of each indicator.
This structured approach enhances evaluation transparency and enables actionable insights for improving dataset quality.

In summary, our contributions are threefold. First, we provide a detailed and reproducible FAIRness assessment of voice biomarker datasets in two clinically significant domains. Second, we propose a novel scoring methodology that improves FAIRness interpretability by accounting for partial compliance and indicator priority. Third, we offer practical, domain-specific recommendations to improve dataset usability, AI-readiness, and clinical integration.

The remainder of the paper is organized as follows. Section~\ref{sec:methodology} describes our methodology, including dataset selection criteria and the structured FAIRness scoring approach. Section~\ref{sec:results} presents and analyzes our findings, highlighting key trends, limitations, and actionable recommendations. Finally, Section~\ref{sec:conclusion} summarizes our contributions and discusses directions for future work.

\section{Methodology}
\label{sec:methodology}

\subsection{Overview of Methodological Approach}

\begin{sloppy}
Our methodology provides a structured and rigorous framework for systematically evaluating the FAIRness of publicly available voice biomarker datasets, focusing on mental health and neurodegenerative conditions. We adopted the well-established FAIR Data Maturity Model~\cite{bahimFAIRDataMaturity2020}, carefully refining it with domain-specific clarifications and a priority-weighted scoring mechanism to enhance interpretability and practical relevance.
\end{sloppy}

\subsection{Dataset Identification and Selection}

We systematically identified candidate datasets using Google Dataset Search with a comprehensive set of queries focused on mental health and neurodegenerative disease voice datasets. The queries included \textit{mental health voice dataset}, \textit{depression voice dataset}, \textit{anxiety speech dataset}, \textit{bipolar disorder speech dataset}, \textit{neurodegenerative disease voice dataset}, \textit{Alzheimer's disease voice dataset}, \textit{Parkinson's disease speech dataset}, and similar variations, ensuring comprehensive coverage of available resources.

\subsubsection{Dataset Inclusion and Exclusion Criteria:}
~Datasets were included if they satisfied the following criteria:
(a) Publicly accessible voice or acoustic datasets.
(b) Targeted mental health conditions (e.g., depression, anxiety, bipolar disorder) or
neurodegenerative conditions (e.g., Alzheimer's, Parkinson's, ALS).
(c) Clearly documented data characteristics and metadata or associated publication describing dataset properties.

Datasets were excluded if they were privately accessible, lacked clear metadata, had ambiguous usage or licensing terms, or were irrelevant to the stated disease categories.

\subsubsection{Final Dataset Selection:}

A total of 27 datasets were selected for our analysis. The selected datasets comprised: \textbf{10 Mental Health datasets}: M1~\cite{baileyGenderBiasDepression2021}, M2~\cite{tlachacEMUEarlyMental2021a}, M3~\cite{ciftciTurkishAudioVisualBipolar}, M4~\cite{AmodMental_health_counseling_conversationsDatasets2024}, M5~\cite{alghifariDevelopmentSorrowAnalysis2023}, M6~\cite{chowdhuryHarnessingLargeLanguage2024}, M7~\cite{islamMentalHealthDiagnosis2024}, M8~\cite{dorahyComparisonAuditoryHallucinations2023}, M9~\cite{yeTemporalModelingMatters2023}, M10~\cite{johnsonBridge2AIVoiceEthicallysourcedDiverse}; and \textbf{17 Neurodegenerative datasets}: N1~\cite{heckerVoiceAnalysisNeurological2022}, N2~\cite{huangAutomaticSpeechAnalysis2024}, N3~\cite{vashkevichBulbarALSDetection2019}, N4~\cite{vashkevichClassificationALSPatients2021}, N5~\cite{littleSuitabilityDysphoniaMeasurements2009}, N6~\cite{hernandezALSDiseasePatient2025}, N7~\cite{baileyGenderBiasDepression2021}, N8~\cite{ballardLogopenicNonfluentVariants2014}, N9~\cite{dubbiosoVoiceSignalsDatabase2024}, N10~\cite{j.hlavnikaEarlyBiomarkersParkinsons2017}, N11~\cite{prezParkinsonDatasetReplicated2016}, N12~\cite{littleExploitingNonlinearRecurrence2007}, N13~\cite{c.sakarParkinsonsDiseaseClassification2018}, N14~\cite{olcaykursunParkinsonsSpeechMultiple2013}, N15~\cite{tsanasAccurateTelemonitoringParkinsons2010}, N16~\cite{ruskoSlovakDatabaseSpeech2024}, N17~\cite{moro-velazquezPhoneticRelevancePhonemic2019}.

\subsection{FAIR Data Maturity Model and Clarifications}

We utilized the FAIR Data Maturity Model~\cite{bahimFAIRDataMaturity2020} to systematically assess dataset FAIRness. The model comprises 15 subprinciples grouped under the four FAIR principles and includes a total of 41 specific assessment indicators. While the original model defines many indicators unambiguously, certain indicators required domain-specific clarifications to ensure consistency and accuracy for voice biomarker datasets. These clarifications are summarized in Table~\ref{tab:fair}; indicators marked as ``(Original definition)'' directly follow the definitions provided by the original FAIR Data Maturity Model without modification.

{\small
\begin{longtable}{|p{2.4cm}|p{1.5cm}|p{6.4cm}|p{1.5cm}|}
\caption{FAIR Indicators with Clarifications and Priority Levels}
\label{tab:fair}\\
\hline
\textbf{FAIR ID} & \textbf{Sub-principle} & \textbf{Clarification (if applicable)} & \textbf{Priority}\\
\hline
\endfirsthead

\hline
\textbf{FAIR ID} & \textbf{Sub-principle} & \textbf{Clarification (if applicable)} & \textbf{Priority}\\
\hline
\endhead

\hline
\multicolumn{4}{|r|}{\textit{Continued on next page}} \\
\hline
\endfoot

\hline
\endlastfoot

\textbf{RDA-F1-01M} & F1 & \textit{Metadata has a DOI or other permanent ID.} & Essential\\
\textbf{RDA-F1-01D} & F1 & \textit{Data has a DOI or other permanent ID.} & Essential\\
\textbf{RDA-F1-02M} & F1 & \textit{Metadata has a DOI or other unique identifier.} & Essential\\
\textbf{RDA-F1-02D} & F1 & \textit{Data has a DOI or other unique identifier.} & Essential\\
\textbf{RDA-F2-01M} & F2 & \textit{Metadata must describe exact data structure or link to a paper detailing the dataset.} & Essential\\
\textbf{RDA-F3-01M} & F3 & \textit{Metadata includes the DOI or some other identifier for the data.} & Essential\\
\textbf{RDA-F4-01M} & F4 & \textit{The metadata is indexed in a search engine or other services (e.g., Google Dataset Search and Kaggle).} & Essential\\
RDA-A1-01M & A1 & (Original definition) & Important\\
RDA-A1-02M & A1 & (Original definition) & Essential\\
RDA-A1-02D & A1 & (Original definition) & Essential\\
\textbf{RDA-A1-03M} & A1 & \textit{Identifier must directly resolve to metadata description page.} & Essential\\
\textbf{RDA-A1-03D} & A1 & \textit{Identifier must directly resolve to data file or download page.} & Essential\\
RDA-A1-04M & A1 & (Original definition) & Essential\\
RDA-A1-04D & A1 & (Original definition) & Essential\\
RDA-A1-05D & A1 & (Original definition) & Important\\
RDA-A1.1-01M & A1.1 & (Original definition) & Essential\\
RDA-A1.1-01D & A1.1 & (Original definition) & Important\\
RDA-A1.2-01D & A1.2 & (Original definition) & Useful\\
RDA-A2-01M & A2 & (Original definition) & Essential\\
RDA-I1-01M & I1 & (Original definition) & Important\\
RDA-I1-01D & I1 & (Original definition) & Important\\
RDA-I1-02M & I1 & (Original definition) & Important\\
RDA-I1-02D & I1 & (Original definition) & Important\\
\textbf{RDA-I2-01M} & I2 & \textit{Metadata must use FAIR-compliant vocabulary or ontology.} & Important\\
\textbf{RDA-I2-01D} & I2 & \textit{Data must use FAIR-compliant vocabulary or ontology.} & Useful\\
RDA-I3-01M & I3 & (Original definition) & Important\\
RDA-I3-01D & I3 & (Original definition) & Useful\\
RDA-I3-02M & I3 & (Original definition) & Useful\\
RDA-I3-02D & I3 & (Original definition) & Important\\
RDA-I3-03M & I3 & (Original definition) & Important\\
RDA-I3-04M & I3 & (Original definition) & Useful\\
\textbf{RDA-R1-01M} & R1 & \textit{Metadata must be complete as described in RDA-F2-01M and fulfill multiple other FAIR indicators.} & Essential\\
RDA-R1.1-01M & R1.1 & (Original definition) & Essential\\
RDA-R1.1-02M & R1.1 & (Original definition) & Important\\
RDA-R1.1-03M & R1.1 & (Original definition) & Important\\
RDA-R1.2-01M & R1.2 & (Original definition) & Useful\\
RDA-R1.2-02M & R1.2 & (Original definition) & Useful\\
RDA-R1.3-01M & R1.3 & (Original definition) & Essential\\
RDA-R1.3-01D & R1.3 & (Original definition) & Essential\\
RDA-R1.3-02M & R1.3 & (Original definition) & Essential\\
RDA-R1.3-02D & R1.3 & (Original definition) & Essential\\
\end{longtable}}

\subsection{Priority-Weighted FAIRness Scoring Methodology}

We evaluated FAIRness at four distinct levels: indicator, subprinciple, principle (F, A, I, R), and composite (FAIR). At the \textit{indicator level}, each FAIR indicator was assessed as a binary variable (satisfied = 1, not satisfied = 0). At the \textit{subprinciple level}, scores were computed as follows:
\[
s_i =
\begin{cases}
0 & \text{if all indicators for subprinciple were not satisfied}\\[6pt]
0.5 & \text{if some indicators for subprinciple were satisfied and others not}\\[6pt]
1 & \text{if all indicators for subprinciple were satisfied}
\end{cases}
\]

To accurately reflect the varying importance of indicators, we employed a priority weighting scheme in which Essential indicators were assigned a weight of 4, Important indicators a weight of 3, and Useful indicators a weight of 1. These weights can be readily adjusted according to end-user preferences or domain-specific considerations, allowing flexibility in FAIRness evaluations.

At the \textit{principle level} (F, A, I, R), continuous scores between 0 and 1 were computed using a weighted average of the relevant subprinciple scores. Similarly, a single \textit{composite FAIRness score} was computed across all principles, also ranging continuously from 0 to 1. Specifically, the weighted FAIR score \(S\), for both the composite and individual principle levels, was calculated in two steps. First, we computed the weighted average priority (\(w_i\)) for each subprinciple \(i\):
\[
w_i = \frac{1}{m_i}\sum_{j=1}^{m_i} w_{ij},
\]
where \( w_{ij} \) represents the weight (Essential = 4, Important = 3, Useful = 1) of the \( j^{th} \) indicator within the \( i^{th} \) subprinciple, and \( m_i \) is the total number of indicators within that subprinciple. The FAIR score (\(S\))—either for a single principle or as a composite across all principles—was then calculated as a weighted average of the corresponding subprinciple scores:
\[
S = \frac{\sum_{i=1}^{n} w_i s_i}{\sum_{i=1}^{n} w_i},
\]
where \( s_i \) is the score of the \( i^{th} \) subprinciple, and \( n \) is the total number of subprinciples considered.

The resulting subprinciple-level, principle-level, and composite FAIRness scores are presented and analyzed for the selected datasets in the next section.

\subsection{Ensuring Evaluation Accuracy and Consistency}

To ensure rigorous and reliable evaluations, we developed and utilized a detailed scoring rubric, minimizing ambiguity and subjectivity. Each dataset was scored systematically following these clearly documented guidelines (see Table~\ref{tab:fair}). This structured approach significantly enhances the transparency, reproducibility, and credibility of our FAIRness assessment, ensuring confidence in our methodology and results.

%\vspace*{-0.1in}
\section{Results and Discussion}
\label{sec:results}

%\vspace*{-0.2in}
\subsection{FAIRness Across Subprinciples, Principles, and Composite Scores}

\begin{figure}[t]
    \centering
    \includegraphics[width=\textwidth]{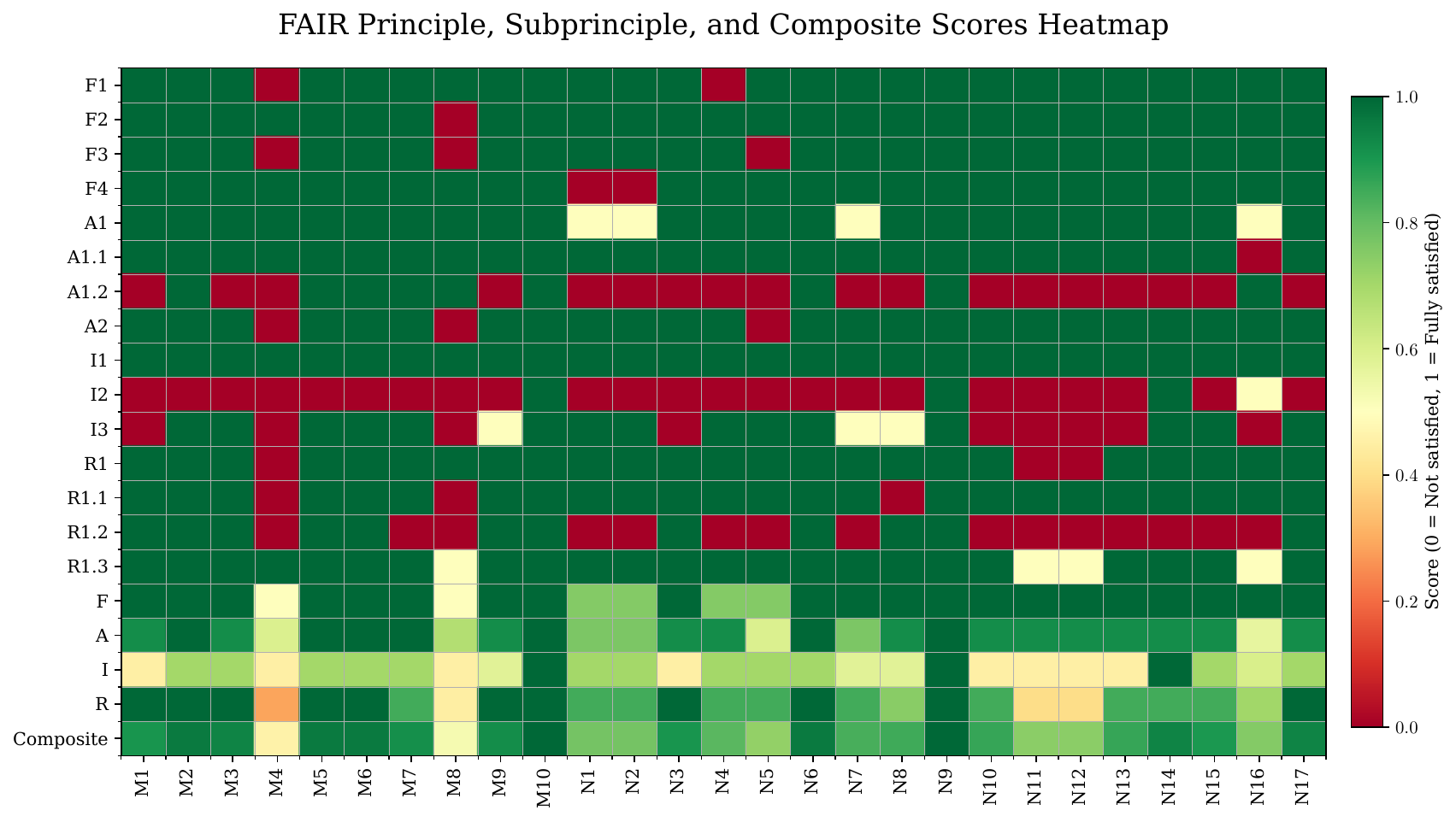}
    \caption{FAIRness heatmap displaying subprinciple, principle, and composite scores per dataset.}
\label{fig:heatmap}
%\vspace*{-0.25in}
\end{figure}

We begin our analysis by evaluating FAIRness at the subprinciple, principle, and composite levels across the 27 selected voice biomarker datasets (Fig.~\ref{fig:heatmap}). The heatmap illustrates scores ranging from 0 (not satisfied) to 1 (fully satisfied) for each dataset, clearly summarizing the comprehensive FAIRness evaluation.

Several clear patterns emerge from Fig.~\ref{fig:heatmap}. Most notably, the majority of datasets consistently exhibited high performance across all four Findability (F) subprinciples (F1–F4). This result indicates broad adoption of fundamental findability practices, including DOI assignment and indexing, enabling robust dataset discoverability. However, substantial variability and weaknesses became apparent within the Accessibility (A), Interoperability (I), and Reusability (R) principles.

In particular, the heatmap reveals frequent low or zero scores within subprinciples A1.2 (standardized protocols for data access), I2 (use of FAIR-compliant vocabularies), I3 (use of standard metadata formats), and R1.2 (availability of provenance metadata). These results demonstrate significant gaps in providing clear access mechanisms, standardized vocabularies, structured metadata documentation, and data provenance details—critical components necessary for dataset interoperability and reuse.

Examining scores at the principle level, Findability consistently outperformed other principles, whereas Accessibility, Interoperability, and Reusability scores showed marked variability and generally lower performance. This trend clearly reflects widespread challenges within the community regarding metadata standardization, structured vocabularies, and accessibility protocols.

The composite FAIRness scores, shown in the bottom row of Fig.~1, summarize overall dataset quality. They clearly illustrate the direct impact of identified gaps, resulting in a broad range of composite scores across datasets. Addressing the identified subprinciple-level weaknesses—in particular within Accessibility, Interoperability, and Reusability—will significantly improve composite FAIRness, thereby enhancing dataset reuse potential and clinical research value.

\subsection{FAIRness by Disease Category}

\begin{figure}[t]
%\vspace*{-0.15in}
    \centering
    \includegraphics[width=\textwidth]{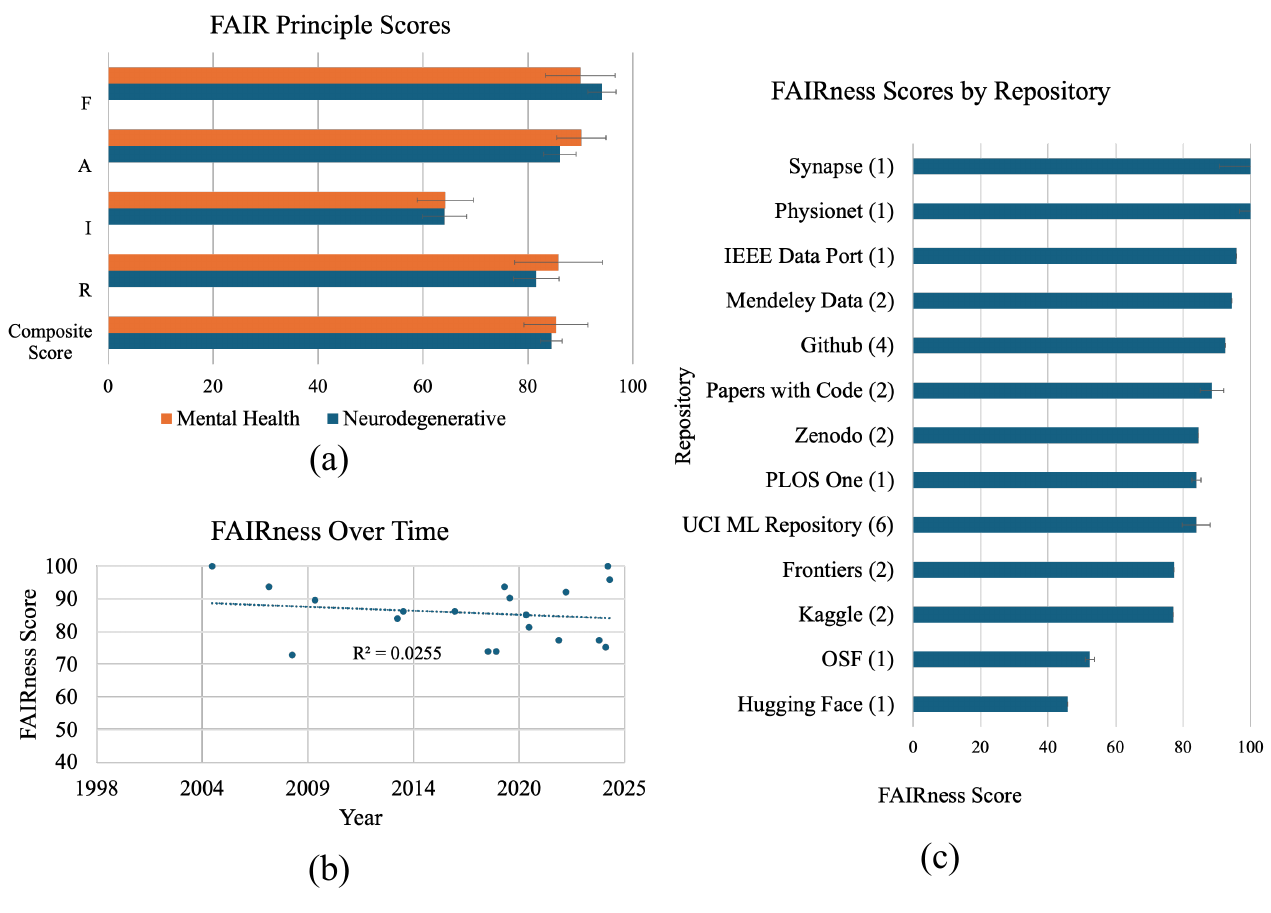}
    %\vspace*{-0.25in}
    \caption{FAIRness results:
    (a) FAIRness principle and composite scores across mental health and neurodegenerative disease categories. 
    (b) FAIRness composite scores over time. 
    (c) FAIRness composite scores by repository.}
\label{fig:fair_composite_results}
%\vspace*{-0.25in}
\end{figure}

We next examined differences in FAIRness across disease categories by comparing FAIR principle (F, A, I, R) and composite scores between mental health and neurodegenerative datasets (Fig.~\ref{fig:fair_composite_results}(a)).

Our analysis revealed nuanced but clear differences. Neurodegenerative datasets slightly outperformed mental health datasets in Findability (F), indicating more consistent adherence to data discoverability standards. Conversely, mental health datasets demonstrated clearly higher scores—though still close to neurodegenerative datasets—in both Accessibility (A) and Reusability (R), suggesting comparatively better adherence to access protocols and documentation standards for dataset reuse.

For Interoperability (I) and composite FAIRness scores, the two disease categories were closely aligned, indicating comparable overall FAIR compliance levels. However, mental health datasets exhibited greater variability across FAIR principles, highlighting inconsistent implementation of FAIR practices in this category compared to the relatively uniform FAIRness scores among neurodegenerative datasets.

These findings emphasize the need for tailored improvement strategies in each domain. Mental health research would benefit from efforts to achieve more consistent adoption of FAIR practices across principles, reducing variability. Conversely, neurodegenerative datasets should specifically enhance their accessibility and reusability documentation. Addressing these targeted domain-specific gaps will improve overall FAIR compliance.

\subsection{Temporal Trends in Dataset FAIRness}

To evaluate temporal trends in dataset FAIRness, we analyzed composite FAIRness scores against the publication years of the 27 selected datasets (Fig.~\ref{fig:fair_composite_results}(b)). Contrary to initial expectations, our analysis revealed no clear improvement trend in FAIRness scores over the time span considered (2004–2025).

The data exhibited considerable variability in composite FAIRness scores across different years, with several datasets achieving high FAIRness scores even in earlier periods. Notably, datasets published in recent years (2020–2025) did not consistently demonstrate higher FAIRness scores compared to earlier datasets. Instead, FAIRness scores showed a slight declining trend, as indicated by the weak negative slope of the regression line and the very low coefficient of determination (R² = 0.0255). This result clearly suggests that the increased awareness of FAIR principles in recent years has not yet resulted in consistent improvements across voice biomarker datasets.

These findings highlight a critical gap between the widespread endorsement of FAIR principles and their actual implementation within publicly available voice biomarker datasets. The observed variability underscores the need for consistent community-wide efforts, standardized guidelines, and improved incentives to systematically integrate FAIR principles into dataset creation and publication practices moving forward.

\subsection{Repository-Based FAIRness Variability}

We examined composite FAIRness scores across different repositories hosting the selected voice biomarker datasets (Fig.~\ref{fig:fair_composite_results}(c)). Our analysis clearly revealed that FAIRness scores varied substantially based on the hosting repository, highlighting the significant impact repository practices and infrastructure have on dataset FAIRness.

The repositories Synapse and Physionet achieved the highest FAIRness scores, each hosting a single dataset that scored exceptionally well. Similarly, IEEE Data Port, Mendeley Data, GitHub, and Papers with Code demonstrated consistently high FAIRness scores, indicating effective repository practices and clear adherence to FAIR principles among datasets hosted there. Notably, GitHub datasets exhibited strong FAIRness performance despite the platform being primarily oriented toward code sharing, suggesting that effective metadata practices and structured documentation significantly enhanced FAIR compliance.

In contrast, repositories such as Kaggle, OSF, and Hugging Face showed markedly lower average FAIRness scores. This result suggests deficiencies in consistent adoption of structured metadata, clear accessibility protocols, or other FAIR-compliant practices within these repositories. Additionally, UCI Machine Learning Repository datasets exhibited intermediate FAIRness scores with some variability, reflecting mixed practices in FAIR implementation among datasets hosted on this popular repository.

The observed variation underscores the critical role that repository choice plays in dataset FAIRness. These results clearly indicate the necessity for dataset creators and maintainers to carefully consider repository selection, favoring platforms with established FAIR-compliant practices and infrastructure. Improving FAIR compliance at the repository level, particularly among lower-performing repositories, will significantly enhance dataset interoperability, reuse, and research impact within voice biomarker research communities.

\subsection{Practical Recommendations}

Our comprehensive FAIRness analysis of publicly available voice biomarker datasets highlights several critical areas for improvement. To address identified gaps, we recommend the following: (a) \textbf{Structured metadata guidelines:} The development and consistent adoption of clear domain-specific metadata standards tailored to voice biomarker research are essential for improving interoperability and reusability.
(b) \textbf{Repository selection:} Researchers should choose repositories with proven FAIR-compliant practices, robust metadata standards, and clear data access policies.
(c) \textbf{Routine FAIR evaluation:} Regular use of structured, priority-weighted FAIR evaluation frameworks (as applied here) is recommended to systematically identify and address dataset weaknesses and enhance overall FAIR compliance.

Adopting these recommendations will substantially improve data reuse potential, reproducibility of voice biomarker research, and facilitate the clinical translation of voice biomarkers.

\section{Conclusion and Future Work}
\label{sec:conclusion}

This study presented the first systematic, priority-weighted FAIRness evaluation specifically targeting publicly available voice biomarker datasets within mental health and neurodegenerative disease research. Using the FAIR Data Maturity Model, we assessed FAIRness at multiple levels across 27 representative datasets, identifying clear strengths in Findability practices but critical gaps in Accessibility, Interoperability, and Reusability. Practical recommendations, including structured metadata adoption, careful repository selection, and routine FAIR assessments, were provided to address these gaps and improve dataset reuse and reproducibility.

Future work will extend this analysis by evaluating FAIRness across additional disease domains, providing broader insights into cross-domain FAIRness trends. Additionally, comparing our structured evaluation results with automated FAIR assessment tools, such as F-UJI and FAIR Evaluator, will help validate and refine FAIRness assessment methodologies, facilitating efficient, large-scale dataset evaluations. Finally, investigating relationships between dataset FAIRness scores and the downstream performance of AI and machine learning models trained on these datasets could strongly reinforce the critical importance of FAIR dataset quality for advancing voice biomarker research into clinical applications.

\begin{credits}

\begin{comment}
\subsubsection{\ackname} A bold run-in heading in small font size at the end of the paper is
used for general acknowledgments, for example: This study was funded
by X (grant number Y).
\end{comment}

\subsubsection{\discintname}
The authors have no competing interests to declare that are
relevant to the content of this article.
\end{credits}
%
% ---- Bibliography ----
%
% BibTeX users should specify bibliography style 'splncs04'.
% References will then be sorted and formatted in the correct style.
%

\begin{comment}
{
\begingroup
\renewcommand{\section}[2]{}% remove references heading
\setlength{\parskip}{0pt}
\setlength{\itemsep}{0pt}
\setlength{\parsep}{0pt}
\linespread{0.5}\selectfont % aggressive line spacing
\tiny % smallest font size allowed
\bibliographystyle{splncs04}
\bibliography{refs_compact}
\endgroup
}
\end{comment}

% Compact bibliography spacing
%\apptocmd{\thebibliography}{\setlength{\itemsep}{-3pt}}{}{}

\bibliographystyle{splncs04}
\bibliography{refs}

\begin{thebibliography}{10}
\providecommand{\url}[1]{\texttt{#1}}
\providecommand{\urlprefix}{URL }
\providecommand{\doi}[1]{https://doi.org/#1}

\bibitem{alghifariDevelopmentSorrowAnalysis2023}
Alghifari, M.F., Gunawan, T.S., Kartiwi, M.: {Development of sorrow analysis dataset for speech depression prediction}. In: IEEE International Instrumentation and Measurement Technology Conference. pp.~1--6. Kuala Lumpur, Malaysia (May 2023)

\bibitem{AmodMental_health_counseling_conversationsDatasets2024}
Amod: {Amod mental-health-counseling conversations dataset}. Hugging Face Datasets (2024)

\bibitem{bahimFAIRDataMaturity2020}
Bahim, C., Casorr{\'a}n-Amilburu, C., Dekkers, M., Herczog, E., Loozen, N., Repanas, K., Russell, K., Stall, S.: {The FAIR data maturity model: An approach to harmonise FAIR assessments}. Data Science Journal  \textbf{19} (Oct 2020)

\bibitem{baileyGenderBiasDepression2021}
Bailey, A., Plumbley, M.D.: {Gender bias in depression detection using audio features}. arXiv preprint (2021)

\bibitem{ballardLogopenicNonfluentVariants2014}
Ballard, K.J., Savage, S., Leyton, C.E., Vogel, A.P., Hornberger, M., Hodges, J.R.: {Logopenic and nonfluent variants of primary progressive aphasia are differentiated by acoustic measures of speech production}. PLOS ONE  \textbf{9}(2) (Feb 2014)

\bibitem{chen2024review}
Chen, S., Li, L., Han, S., Luo, W., Wang, W., Yang, Y., Wang, X., Zhang, W., Chen, M., Wang, Z.: {Review of voice biomarkers in the screening of neurodegenerative diseases}. Interdisciplinary Nursing Research  \textbf{3}(3) (2024)

\bibitem{chowdhuryHarnessingLargeLanguage2024}
Chowdhury, A.K., Sujon, S.R., Shafi, M.S.S., Ahmmad, T., Ahmed, S., Hasib, K.M., Shah, F.M.: {Harnessing large language models over transformer models for detecting Bengali depressive social media text: A comprehensive study}. Natural Language Processing Journal  \textbf{7} (2024)

\bibitem{ciftciTurkishAudioVisualBipolar}
{\c{C}}ift{\c{c}}i, E., Kaya, H., G{\"u}le{\c{c}}, H., Salah, A.A.: The {Turkish} audio-visual bipolar disorder corpus. In: 2018 First Asian Conference on Affective Computing and Intelligent Interaction (ACII Asia). pp.~1--6. IEEE (2018)

\bibitem{clarkAIreadinessBiomedicalData2024}
Clark, T., Caufield, H., Parker, J.A., Al~Manir, S., Amorim, E., Eddy, J., Gim, N., Gow, B., Goar, W., Haendel, M., et~al.: Ai-readiness for biomedical data: Bridge2ai recommendations. bioRxiv  (2024)

\bibitem{cummins2018speech}
Cummins, N., Baird, A., Schuller, B.W.: {Speech analysis for health: Current state-of-the-art and the increasing impact of deep learning}. Methods  \textbf{151} (2018)

\bibitem{devarajuAutomatedSolutionMeasuring2021}
Devaraju, A., Huber, R.: {An automated solution for measuring the progress toward FAIR research data}. Patterns  \textbf{2}(11) (Nov 2021)

\bibitem{dorahyComparisonAuditoryHallucinations2023}
Dorahy, M.J., Nesbit, A., Palmer, R., Wiltshire, B., Cording, J.R., Hanna, D., Seager, L., Middleton, W.: {A comparison between auditory hallucinations, interpretation of voices, and formal thought disorder in dissociative identity disorder and schizophrenia spectrum disorders}. Journal of Clinical Psychology  \textbf{79}(9) (2023)

\bibitem{dubbiosoVoiceSignalsDatabase2024}
Dubbioso, R., Spisto, M., Verde, L., Iuzzolino, V.V., Senerchia, G., Salvatore, E., De~Pietro, G., De~Falco, I., Sannino, G.: {Voice signals database of ALS patients with different dysarthria severity and healthy controls}. Scientific Data  \textbf{11}(1) (Jul 2024)

\bibitem{fagherazziVoiceHealthUse2021}
Fagherazzi, G., Fischer, A., Ismael, M., Despotovic, V.: {Voice for health: The use of vocal biomarkers from research to clinical practice}. Digital Biomarkers  \textbf{5}(1) (Apr 2021)

\bibitem{gaignard2023fair}
Gaignard, A., Rosnet, T., de~Lamotte, F., Lefort, V., Devignes, M.D.: {FAIR-checker: Supporting digital resource findability and reuse with knowledge graphs and semantic web standards}. Journal of Biomedical Semantics  \textbf{14}(1) (2023)

\bibitem{groth2020fair}
Groth, P., Cousijn, H., Clark, T., Goble, C.: {FAIR data reuse—The path through data citation}. Data Intelligence  \textbf{2}(1–2) (2020)

\bibitem{heckerVoiceAnalysisNeurological2022}
Hecker, P., Steckhan, N., Eyben, F., Schuller, B.W., Arnrich, B.: {Voice analysis for neurological disorder recognition—A systematic review and perspective on emerging trends}. Frontiers in Digital Health  \textbf{4} (Jul 2022)

\bibitem{hernandezALSDiseasePatient2025}
Hernandez, J.P.: {ALS disease patient classification}. Mendeley Data  \textbf{2} (Feb 2025)

\bibitem{j.hlavnikaEarlyBiomarkersParkinsons2017}
Hlavnika, J., Tykalov, T.: {Early biomarkers of Parkinson's disease based on natural connected speech}. UCI Machine Learning Repository (2017)

\bibitem{huangAutomaticSpeechAnalysis2024}
Huang, L., Yang, H., Che, Y., Yang, J.: {Automatic speech analysis for detecting cognitive decline of older adults}. Frontiers in Public Health  \textbf{12} (Aug 2024)

\bibitem{islamMentalHealthDiagnosis2024}
Islam, R., Ahad, M.T., Ahmed, F., Song, B., Li, Y.: {Mental health diagnosis from voice data using convolutional neural networks and vision transformers}. Journal of Voice  (Nov 2024)

\bibitem{johnsonBridge2AIVoiceEthicallysourcedDiverse}
Johnson, A., Bélisle-Pipon, J.C., Dorr, D., Ghosh, S., et~al.: {Bridge2AI-Voice: An ethically-sourced, diverse voice dataset linked to health information}. PhysioNet

\bibitem{littleSuitabilityDysphoniaMeasurements2009}
Little, M.A., McSharry, P.E., Hunter, E.J., Spielman, J., Ramig, L.O.: {Suitability of dysphonia measurements for telemonitoring of Parkinson's disease}. IEEE Transactions on Biomedical Engineering  \textbf{56}(4) (Apr 2009)

\bibitem{littleExploitingNonlinearRecurrence2007}
Little, M.A., McSharry, P.E., Roberts, S.J., Costello, D.A.E., Moroz, I.M.: {Exploiting nonlinear recurrence and fractal scaling properties for voice disorder detection}. BioMedical Engineering OnLine  \textbf{6}(1) (Jun 2007)

\bibitem{low2020automated}
Low, D.M., Bentley, K.H., Ghosh, S.S.: {Automated assessment of psychiatric disorders using speech: A systematic review}. Laryngoscope Investigative Otolaryngology  \textbf{5}(1) (2020)

\bibitem{moro-velazquezPhoneticRelevancePhonemic2019}
Moro-Velazquez, L., Gomez-Garcia, J.A., Godino-Llorente, J.I., Grandas-Perez, F., Shattuck-Hufnagel, S., Yag{\"u}e-Jimenez, V., Dehak, N.: {Phonetic relevance and phonemic grouping of speech in the automatic detection of Parkinson's disease}. Scientific Reports  \textbf{9}(1) (Dec 2019)

\bibitem{olcaykursunParkinsonsSpeechMultiple2013}
Olcay~Kursun, B.S.: {Parkinson's speech with multiple types of sound recordings}. UCI Machine Learning Repository (2013)

\bibitem{petrosyanFAIRDegreeAssessment2023}
Petrosyan, L., Aleixandre-Benavent, R., Peset, F., Valderrama-Zuri{\'a}n, J.C., Ferrer-Sapena, A., Sixto-Costoya, A.: {FAIR degree assessment in agriculture datasets using the F-UJI tool}. Ecological Informatics  \textbf{76} (Sep 2023)

\bibitem{prezParkinsonDatasetReplicated2016}
Prez, C.: {Parkinson dataset with replicated acoustic features}. UCI Machine Learning Repository (2016)

\bibitem{robin2020evaluation}
Robin, J., Harrison, J.E., Kaufman, L.D., Rudzicz, F., Simpson, W., Yancheva, M.: {Evaluation of speech-based digital biomarkers: Review and recommendations}. Digital Biomarkers  \textbf{4}(3) (2020)

\bibitem{ruskoSlovakDatabaseSpeech2024}
Rusko, M., Sabo, R., Trnka, M., Zimmermann, A., Malaschitz, R., Ru{\v z}ick{\'y}, E., Brandoburov{\'a}, P., Kevick{\'a}, V., {\v S}korv{\'a}nek, M.: {Slovak database of speech affected by neurodegenerative diseases}. Scientific Data  \textbf{11}(1) (Dec 2024)

\bibitem{c.sakarParkinsonsDiseaseClassification2018}
Sakar, G.S.C.: {Parkinson's disease classification}. UCI Machine Learning Repository (2018)

\bibitem{stellmachHowAssessFAIRness2024}
Stellmach, C., Muzoora, M.R.: {How to assess FAIRness of your data—A summary of testing two FAIR validators}. In: MEDINFO 2023 – The Future Is Accessible. pp. 154--158. Sydney, Australia (Jan 2024)

\bibitem{tlachacEMUEarlyMental2021a}
Tlachac, M.L., Toto, E., Lovering, J., Kayastha, R., Taurich, N., Rundensteiner, E.: {EMU: Early mental health uncovering framework and dataset}. In: IEEE International Conference on Machine Learning and Applications. pp. 1311--1318. Pasadena, CA, USA (Dec 2021)

\bibitem{tsanasAccurateTelemonitoringParkinsons2010}
Tsanas, A., Little, M.A., McSharry, P.E., Ramig, L.O.: {Accurate telemonitoring of Parkinson's disease progression by noninvasive speech tests}. IEEE Transactions on Biomedical Engineering  \textbf{57}(4) (Apr 2010)

\bibitem{vashkevichBulbarALSDetection2019}
Vashkevich, M., Petrovsky, A., Rushkevich, Y.: {Bulbar ALS detection based on analysis of voice perturbation and vibrato}. In: Signal Processing: Algorithms, Architectures, Arrangements and Applications. pp. 267--272. Poznań, Poland (Sep 2019)

\bibitem{vashkevichClassificationALSPatients2021}
Vashkevich, M., Rushkevich, Y.: {Classification of ALS patients based on acoustic analysis of sustained vowel phonations}. Biomedical Signal Processing and Control  \textbf{65} (Mar 2021)

\bibitem{wilkinson2016fair}
Wilkinson, M.D., Dumontier, M., Aalbersberg, I.J., Appleton, G., Axton, M., Baak, A., Blomberg, N., Boiten, J.W., da~Silva~Santos, L.B., Bourne, P.E., et~al.: {The FAIR guiding principles for scientific data management and stewardship}. Scientific Data  \textbf{3}(1) (2016)

\bibitem{wilkinson2019evaluating}
Wilkinson, M.D., Dumontier, M., Sansone, S.A., Bonino~da Silva~Santos, L.O., Prieto, M., Batista, D., McQuilton, P., Kuhn, T., Rocca-Serra, P., Crosas, M., et~al.: {Evaluating FAIR maturity through a scalable, automated, community-governed framework}. Scientific Data  \textbf{6}(1) (2019)

\bibitem{yeTemporalModelingMatters2023}
Ye, J., Wen, X.C., Wei, Y., Xu, Y., Liu, K., Shan, H.: {Temporal modeling matters: A novel temporal emotional modeling approach for speech emotion recognition}. In: IEEE International Conference on Acoustics, Speech and Signal Processing. pp.~1--5. Rhodes, Greece (Jun 2023)

\end{thebibliography}

\end{document}